# O(D,D) gauge fields in the T-dual string Lagrangian

Machiko Hatsuda

*Physics Division, Faculty of Medicine, Juntendo University*
*Chiba 270-1695, Japan*
*KEK Theory Center, High Energy Accelerator Research Organization*
*Tsukuba, Ibaraki 305-0801, Japan*

and

Warren Siegel

C. N. Yang Institute for Theoretical Physics
*State University of New York, Stony Brook, NY 11794-3840*

January 9, 2019

**Abstract**

We present the string Lagrangian with manifest T-duality. Not only zero-modes but also all string modes are doubled. The gravitational field is an O(D,D) gauge field. We give a Lagrangian version of the section condition for the gauge invariance which compensates the O(D,D) transformation from the gravitational field and the GL(2D) coordinate transformation. We also show the gauge invariance of the line element of the manifest T-duality space and the O(D,D) condition on the background. Different sections describe dual spaces.

# Contents

# 1 Introduction

T-duality is one of the characteristic features of the low energy effective gravity theory of string. T-duality is a duality between long distance and short distance physics, leading to the existence of a minimal length of the theory. In the long distance regime momentum excitations are dominant, while in the short distance regime winding-mode excitations become dominant. To clarify the minimal length physics a theory is desirable in which both momenta and winding modes are treated equally. Doubling the coordinates corresponding to momenta and winding modes makes T-duality manifest. Duff firstly introduced doubled coordinates to write down duality between the field equation and the Bianchi identity [1]. Tseytlin used doubled coordinates to write down the action with manifest T-duality without worldsheet covariance [2, 3]. Duff and Tseytlin described O(d,d) scalars for d compactified dimensions, with coordinates doubled, in the string Lagrangian. One of the present authors (W.S.) described O(D,D) gauge fields for whole D-dimensional spacetime, with only zero-modes doubled (not all string modes $X(\sigma)$), in the Hamiltonian formalism [4–6]. This manifest T-duality theory (T-theory) is the O(D,D) gauge theory for gravity. A closed string provides the $B$ field as well as the gravitational metric, and they are parameters of the coset O(D,D)/O(D−1,1)$^2$ where O(D−1,1)$^2$ is doubled Lorentz groups. It was further developed for Poincaré systems [7], type II supersymmetric systems [8–11] and a nonableian system [12]. All string modes are doubled in flat space [11]. Here we describe O(D,D) gauge fields in the Lagrangian formalism with doubled coordinates. Massless field theory with O(D,D) T-duality group is called "double field theory" [13] and it now becomes an active area whose review papers are for example [14–17].

An important question is how to reduce the half of the doubled coordinates. So far imposing the selfduality condition on doubled coordinates reduces half of the doubled coordinates or half of the worldsheet chiral currents. Instead of solving the selfduality condition gauging the antiselfdual mode allows to preserve both target space doubling and the worldsheet covariance [18]. In this paper we have applied this formulation to the doubled coordinate space. These covariance are preferable for the quantum computation. The obtained new T-theory Lagrangian with the O(D,D) gravitational fields $E_M{}^A$ is

$$L = e^{-1}(J_+{}^{\bar{A}} J_{-\bar{A}} + J_+{}^{\underline{A}} J_{-\underline{A}} - \hat{\lambda}_+ J_-{}^{\bar{A}} J_{-\bar{A}} - \hat{\lambda}_- J_+{}^{\underline{A}} J_{+\underline{A}}) \qquad (1.1)$$

The Lagrangian includes both the selfdual current $J_+{}^A$ and the antiselfdual current $J_-{}^A$. The current has the worldsheet index and the target space index

$$J_a{}^A \equiv (e_a{}^m \partial_m X^M) E_M{}^A \qquad (1.2)$$

The worldsheet index is $m = (\tau, \sigma)$ and the worldsheet tangent-space index is $a = \pm$. The 2D-dimensional target space index is $M$ and the target tangent-space index is $A = (\bar{A}, \underline{A})$ with left/right indices $\bar{A}, \underline{A}$. The worldsheet zweibein is $e_a{}^m$ with $e = \det e_a{}^m$. The non-chiral treatment [18] has some advantages in the worldsheet covariance comparing to the chiral treatment: the Lagrangian has the manifest 2-dimensional Lorentz symmetry and the canonical $\sigma$ derivative coincides with the chain rule $\sigma$ derivative without the selfduality condition.

Under the general coordinate transformation $X^M \to X'^M = X^M - \Lambda^M$ with the infinitesimal parameter $\Lambda^M(X)$, the gauge invariance of the action requires the Lagrangian version of the section condition

$$\partial_m X^M \eta_{ML} \partial_N \Lambda^L = 0 \tag{1.3}$$

where $\eta_{MN}$ is the O(D,D) invariant metric. Multiplying $d\sigma^m dX^N$ on (1.3) leads to the gauge invariance condition of the orthogonality condition

$$dX^M \eta_{MN} dX^N = dX'^M \eta_{MN} dX'^N \tag{1.4}$$

The new T-theory Hamiltonian includes the antiselfdual currents $\widetilde{\triangleright}_M$, so its gauge invariance requires the Lagrangian version of the section condition and the weaker form of the selfduality condition

$$\partial_\sigma X^M \eta_{ML} \partial_N \Lambda^L = 0 \,, \quad \widetilde{\triangleright}_M \eta^{MN} \partial_N \Lambda^L = 0 \tag{1.5}$$

T-duality is a duality between long and short distances in D-dimensional spacetime, so the distance of the T-theory spacetime is double-faced. Invariance under the general coordinate transformation is the guiding principle to determine the line element of the manifest T-duality space (T-space). We propose the line element of the T-space and the orthogonal condition

$$\begin{aligned} ds^2 &= dX^M G_{MN}(X) dX^N \\ 0 &= dX^M \eta_{MN} dX^N \end{aligned} \tag{1.6}$$

They are fundamental quantities of the stringy geometry. Their general coordinate invariances require the constraint (1.3) which compensate the O(D,D) transformation from the gravitational fields and the GL(2D) transformation from the coordinates. The general coordinate invariance of the orthogonal condition, the second equation in (1.6) or the equation (1.4), leads to the orthogonal condition on the background as $E_A{}^M \eta_{MN} E_B{}^N = \eta_{AB}$. Different sections of the metric (1.6) give dual space metrics.

The organization of the paper is the following: In the next section the T-theory Lagrangian is proposed. The Lagrangian includes both the selfdual and the antiselfdual currents. Selfduality is chirality, where the selfdual current is the chiral left moving current. Contrast to the previous Hamiltonian formulation which contains only selfdual currents [11], the new T-theory Lagrangian is non-chiral. The worldsheet covariance is manifest where the Weyl-Lorentz gauge parameters are used for the worldsheet reparametrization invariances [19]. In section 3 the gauge invariances of the T-theory Hamiltonian and the Lagrangian under the spacetime coordinate transformation are shown.

The spacetime coordinate invariance of the Hamiltonian requires two types of the conditions; the Lagrangian version of the section condition and the weaker form of the selfduality condition. The spacetime coordinate invariance of the Lagrangian requires the new condition. In section 4 constraint algebras of the Virasoro constraints and selfduality constraints are presented. Subsidiary conditions are clarified. The Hamiltonian is non-chiral including both the covariant derivative and the symmetry generator. Non-chiral description makes the worldsheet covariance manifest. The comparison with the chiral description in our previous work [11] is also explained. In section 5 the line element of the T-space and the orthogonal condition on the background are derived from the Virasoro operators. Although the expression of the line element was used to examine solutions [20], the gauge invariance was not examined yet. The O(D,D) gauge invariance of the line element was tried to be realized by integrating out a compensating field [21]. The selfduality condition eliminates $\partial_\sigma X$ by $\partial_\tau X$ resulting the zero-mode limit (particle limit) of the Virasoro operators. Their gauge invariance requires the new condition. The covariant and contravariant vectors are transformed with O(D,D) matrices. This is different from the bilinear of the GL(2D) matrices proposed in [22] which is inconsistent under the multiple transformations and non-associative. The finite O(D,D) transformations are not included there. The correspondence between the Buscher's T-duality transformation and the sectioning in the T-space is also shown by taking an AdS space as a simple example.

## 2 Lagrangian

The Hamiltonian form of the Lagrangian for a bosonic string in a flat space is written by

$$L = \dot{X} \cdot P - H \,, \quad H = g_-\frac{1}{4}(P+X')^2 + g_+\frac{1}{4}(P-X')^2 \tag{2.1}$$

where vectors are contracted by the usual D-dimensional Minkowski metric. We use the simple notation $\dot{X} = \partial_\tau X$ and $X' = \partial_\sigma X$ only for obvious situations. It reduces to

$$L = [\tfrac{1}{2}(g_+ + g_-)]^{-1}\tfrac{1}{2}(\dot{X} + g_+ X')(\dot{X} - g_- X') = \tfrac{1}{2}e^{-1}(e_+ X)(e_- X) \tag{2.2}$$

upon eliminating $P$ by its equation of motion. This suggests the Weyl-Lorentz gauge for the zweibein [19]

$$e_\pm \equiv e_\pm{}^m \partial_m = \partial_\tau \pm g_\pm \partial_\sigma \tag{2.3}$$

or more explicitly

$$e_a{}^m = \begin{matrix} \\ - \\ + \end{matrix}\begin{pmatrix} \tau & \sigma \\ 1 & -g_- \\ 1 & g_+ \end{pmatrix} \tag{2.4}$$

To derive the manifestly covariant Lagrangian for T-theory from the Hamiltonian, it's useful to note the orthogonality of the background field $E_A{}^M$ with respect to the metric $\eta$,

$$E_A{}^M E_B{}^N \eta^{AB} = \eta^{MN}, \quad E_A{}^M E_B{}^N \hat{\eta}^{AB} = G^{MN}, \quad E_A{}^M E_M{}^B = \delta_A^B \tag{2.5}$$

Indices are raised and lowered with $\eta^{MN}$ and $\eta_{MN}$. The double space metric field $G_{MN}$ includes the $B$ field. The doubled (left/right) indices $A = (\bar{A}, \underline{A})$ can be covariantly divided with respect to the tangent-space symmetry, with

$$\eta_{AB} = \begin{pmatrix} \eta_{\bar{A}\bar{B}} & 0 \\ 0 & -\eta_{\underline{AB}} \end{pmatrix}, \quad \hat{\eta}_{AB} = \begin{pmatrix} \eta_{\bar{A}\bar{B}} & 0 \\ 0 & \eta_{\underline{AB}} \end{pmatrix} \tag{2.6}$$

Thus $E_M{}^{\bar{A}} E_{\bar{A}}{}^N$ and $E_M{}^{\underline{A}} E_{\underline{A}}{}^N$ are projection operators, so the result in a background follows directly from that without. (The background can be treated effectively as constants for purposes of varying just the $P$'s.)

The Hamiltonian in the flat doubled space can thus be written as

$$H = g_- \tfrac{1}{4}(\bar{P} + \bar{X}')^2 + (g_+ + \lambda_+) \tfrac{1}{4}(\bar{P} - \bar{X}')^2 + (g_- + \lambda_-) \tfrac{1}{4}(\underline{P} + \underline{X}')^2 + g_+ \tfrac{1}{4}(\underline{P} - \underline{X}')^2 \tag{2.7}$$

The left/right D-dimensional vectors are contracted by the left/right D-dimensional Minkowski metrics as $\bar{V}^2 = \bar{V}_{\bar{M}} \eta^{\overline{MN}} \bar{V}_{\bar{N}}$ and $\underline{V}^2 = \underline{V}_{\underline{M}} \eta^{\underline{MN}} \underline{V}_{\underline{N}}$. The $g$'s play the same role as before for both the $\bar{A}$ (left) and $\underline{A}$ (right) pieces of $X$ and $P$ denoted by $(\bar{X}, \bar{P})$ (left) and $(\underline{X}, \underline{P})$ (right). The $\lambda$'s impose selfduality by killing the antiselfdual pieces, which carry opposite signs for $\bar{A}$ and $\underline{A}$. In terms of O(D,D) coordinates $X^M$ and O(D,D) invariant metric $\eta^{MN}$ the covariant derivative $\overset{\square}{\triangleright}_M$ and the symmetry generator $\widetilde{\triangleright}_M$ are given by

$$\begin{cases} \overset{\square}{\triangleright}_M = P_M + \partial_\sigma X^N \eta_{NM} \\ \widetilde{\triangleright}_M = P_M - \partial_\sigma X^N \eta_{NM} \end{cases} \tag{2.8}$$

The Hamiltonian in (2.7) is rewritten as

$$H = g_- \tfrac{1}{4}(\overset{\square}{\triangleright}_{\bar{M}})^2 + (g_+ + \lambda_+) \tfrac{1}{4}(\widetilde{\triangleright}_{\bar{M}})^2 + (g_- + \lambda_-) \tfrac{1}{4}(\widetilde{\triangleright}_{\underline{M}})^2 + g_+ \tfrac{1}{4}(\overset{\square}{\triangleright}_{\underline{M}})^2 \tag{2.9}$$

In curved space the vielbein is included as $\triangleright_A = E_A{}^M \overset{\square}{\triangleright}_M$ and $\widetilde{\triangleright}_A = E_A{}^M \widetilde{\triangleright}_M$, and the Hamiltonian becomes

$$H = g_- \tfrac{1}{4}(\triangleright_{\bar{A}})^2 + (g_+ + \lambda_+) \tfrac{1}{4}(\widetilde{\triangleright}_{\bar{A}})^2 + (g_- + \lambda_-) \tfrac{1}{4}(\widetilde{\triangleright}_{\underline{A}})^2 + g_+ \tfrac{1}{4}(\triangleright_{\underline{A}})^2 \tag{2.10}$$

The derivation of the Lagrangian is then a double copy of the usual string case, with the appropriate substitutions for the $g$'s. This suggests using independent zweibeins for left and right

$$\begin{pmatrix} \bar{e}_+ & \bar{e}_- \\ \underline{e}_+ & \underline{e}_- \end{pmatrix} = \begin{pmatrix} 1 & 1 \\ 1 & 1 \end{pmatrix} \partial_\tau + \begin{pmatrix} g_+ + \lambda_+ & -g_- \\ g_+ & -g_- - \lambda_- \end{pmatrix} \partial_\sigma \tag{2.11}$$

The final result for the T-theory Lagrangian is then

$$L = \tfrac{1}{2}\bar{e}^{-1}\bar{J}_+{}^{\bar{A}}\bar{J}_{-\bar{A}} + \tfrac{1}{2}\underline{e}^{-1}\underline{J}_+{}^{\underline{A}}\underline{J}_{-\underline{A}} \tag{2.12}$$

where the currents with background are

$$\bar{J}_{\bar{a}}{}^{\bar{A}} \equiv (\bar{e}_{\bar{a}}{}^m \partial_m X^M) E_M{}^{\bar{A}}, \quad \underline{J}_{\underline{a}}{}^{\underline{A}} \equiv (\underline{e}_{\underline{a}}{}^m \partial_m X^M) E_M{}^{\underline{A}} \tag{2.13}$$

with $\bar{a} = (+,-) = \underline{a}$, $\bar{e} = \det \bar{e}_{\bar{a}}{}^m$ and $\underline{e} = \det \underline{e}_{\underline{a}}{}^m$. (We have resisted using explicit $\bar{\phantom{x}}$'s and $\underline{\phantom{x}}$'s for $+$ and $-$ indices.) Notation for indices are summarized in the appendix.

The full D+D spacetime coordinate transformations act on the $M$ indices, while two independent local O(D−1,1) tangent-spacetime transformations act on indices $\bar{A}$ and $\underline{A}$. Similarly, the full worldsheet coordinate transformations act on index $m$, while four independent local scale (actually Weyl $\pm$ Lorentz) tangent-worldsheet transformations act on the $+$ and $-$ components of the $\bar{a}$ and $\underline{a}$ indices. Thus there are two independent worldsheet conformal metrics.

Another interesting gauge for half these Weyl/Lorentz invariances is

$$\begin{pmatrix} \bar{e}_+ & \bar{e}_- \\ \underline{e}_+ & \underline{e}_- \end{pmatrix} = \begin{pmatrix} e_+ - \hat{\lambda}_+ e_- & e_- \\ e_+ & e_- - \hat{\lambda}_- e_+ \end{pmatrix} \tag{2.14}$$

Then in terms of the "usual" currents that use only a single zweibein

$$J_a{}^A \equiv (e_a{}^m \partial_m X^M) E_M{}^A \tag{2.15}$$

the above T-theory Lagrangian becomes

$$L = \tfrac{1}{2} e^{-1}(J_+{}^{\bar{A}} J_{-\bar{A}} + J_+{}^{\underline{A}} J_{-\underline{A}} - \hat{\lambda}_+ J_-{}^{\bar{A}} J_{-\bar{A}} - \hat{\lambda}_- J_+{}^{\underline{A}} J_{+\underline{A}}) \tag{2.16}$$

which, directly in the Lagrangian formalism, shows the $\hat{\lambda}$'s as selfduality multipliers. These $\hat{\lambda}$'s are related to the previous by

$$\lambda_\pm = (g_+ + g_-) \frac{\hat{\lambda}_\pm}{1 - \hat{\lambda}_\pm} \tag{2.17}$$

as seen by Weyl/Lorentz rescaling all $\partial_\tau$ coefficients to 1.

## 3 Spacetime coordinate invariance

We show the spacetime coordinate invariances of the T-theory Hamiltonian and Lagrangian requiring new conditions. The transformation of the spacetime background vielbein under $X^M \to X'^M = X^M - \Lambda^M$ was already given as the new Lie derivative in the Hamiltonian formalism [4,5]:

$$\mathcal{L} E_A{}^M = \Lambda^N \partial_N E_A{}^M - E_A{}^N \partial_N \Lambda^M + E_A{}^N \partial^M \Lambda_N \tag{3.1}$$

The only difference from the usual Lie derivative comes from the last term which follows $E_A{}^M$ to be orthogonal.

At first we examine the spacetime coordinate invariance of the T-theory Hamiltonian (2.10). The general coordinate transformation rule is obtained by taking commutator with $-i \int \Lambda^M \overset{\square}{\triangleright}_M$;

$$\delta X^M = -\Lambda^M \ , \ \delta P_M = \partial_M \Lambda^N P_N + \partial_{[M} \Lambda_{N]} \partial_\sigma X^N \tag{3.2}$$

An action should be invariant under combined active and passive transformations, so we combine the new Lie derivative (3.1) and (3.2). Spacetime coordinate transformations of the covariant derivative $\triangleright_A$ and the symmetry generator $\widetilde{\triangleright}_A$ are

$$\begin{aligned}
\delta \triangleright_A + (\mathcal{L} E_A{}^M) \overset{\square}{\triangleright}_M &= E_A{}^N \widetilde{\triangleright}^M \partial_M \Lambda_N \\
\delta \widetilde{\triangleright}_A + (\mathcal{L} E_A{}^M) \widetilde{\triangleright}_M &= E_A{}^N \left( \widetilde{\triangleright}^M \partial_M \Lambda_N + 2 \partial_\sigma X^M \partial_N \Lambda_M \right)
\end{aligned} \tag{3.3}$$

Invariance of $\triangleright_A$ and $\widetilde{\triangleright}_A$ requires the weaker form of the selfduality condition and the Lagrangian version of the section condition

$$\widetilde{\triangleright}^M \partial_M \Lambda_N = 0 \ , \ \partial_\sigma X^M \partial_N \Lambda_M = 0 \tag{3.4}$$

The fact that vielbein fields $E_A{}^N(X)$ and $E'_A{}^M(X')$ are elements of O(D,D) leads to that the transformation matrix $M_N{}^M$ should be also an O(D,D) element,

$$E'_A{}^M(X') = E_A{}^N(X) M_N{}^M \tag{3.5}$$

The infinitesimal transformation case of $M_N{}^M$ is given by (3.1) as

$$M_N{}^M = \delta_N^M - \partial_N \Lambda^M + \partial^M \Lambda_N \tag{3.6}$$

On the other hand the covariant derivative is transformed

$$\overset{\square}{\triangleright}'_M(X') = N_M{}^N \overset{\square}{\triangleright}_N(X) \tag{3.7}$$

with $\partial_\sigma \Lambda^M = \frac{1}{2}(\overset{\square}{\triangleright}^N - \widetilde{\triangleright}^N) \partial_N \Lambda^M$ and the weaker form of the selfduality constraint in (3.4). The infinitesimal transformation case of $N_M{}^N$ is an O(D,D) element by (3.2) as

$$N_M{}^N = \delta_M^N + \partial_M \Lambda^N - \partial^N \Lambda_M \tag{3.8}$$

Gauge invariance of $\triangleright_A$, (3.3)=0 with the constraints (3.4), leads to the relation between $M_M{}^N$ and $N_M{}^N$ as

$$E_A{}^M(X) \overset{\square}{\triangleright}_M(X) = E'_A{}^M(X') \overset{\square}{\triangleright}'_M(X') \Rightarrow M_N{}^M N_M{}^L = \delta_N^L \tag{3.9}$$

Next we examine the spacetime coordinate invariance of the T-theory Lagrangian. In Lagrangian formulation $J_m^A$ should be also gauge invariant

$$\partial_m X^M E_M{}^A(X) = \partial_m X'^M E'_M{}^A(X') \tag{3.10}$$

The transformation matrices are given by

$$E'_M{}^A(X') = (M^{-1})_M{}^N E_N{}^A(X) \Rightarrow \partial_m X'^M = \partial_m X^N (N^{-1})_N{}^M \tag{3.11}$$

On the other hand the usual chain rule of the derivative gives

$$\partial_m X'^M = \partial_m X^M - \partial_m \Lambda^M = \partial_m X^N (\delta_N^M - \partial_N \Lambda^M) \tag{3.12}$$

These require the following condition

$$\partial_m X^N \partial_M \Lambda_N = 0 \tag{3.13}$$

This condition is the worldsheet covariant version of the last conditions in (3.4).

## 4 Virasoro and selfduality constraints

The manifestly T-dual space is defined by the covariant derivative $\overset{\square}{\triangleright}_M$ in (2.8), while the symmetry generator $\widetilde{\triangleright}$ in (2.8) commute with the covariant derivative. They satisfy

$$\begin{aligned}
[\overset{\square}{\triangleright}_M(1), \overset{\square}{\triangleright}_N(2)] &= -2i\eta_{MN} \partial_\sigma \delta(2-1) \\
[\widetilde{\triangleright}_M(1), \widetilde{\triangleright}_N(2)] &= 2i\eta_{MN} \partial_\sigma \delta(2-1) \\
[\overset{\square}{\triangleright}_M(1), \widetilde{\triangleright}_N(2)] &= 0
\end{aligned} \tag{4.1}$$

with $\partial_\sigma \delta(2-1) = \frac{\partial}{\partial \sigma_2}\delta(\sigma_2 - \sigma_1)$. There is a normalization ambiguity of the currents in (2.8). The commutator of derivative currents gives to the particle derivative as

$$[\overset{\square}{\triangleright}_M, \Phi] = [\widetilde{\triangleright}_M, \Phi] = \frac{1}{i}\partial_M \Phi(X) \tag{4.2}$$

The selfduality condition $\partial_m x = \varepsilon_m{}^n \partial_n \tilde{x}$ is generalized to $\partial_m X^N G_{NM} = \varepsilon_m{}^n \partial_n X^N \eta_{NM}$ with $X^M = (x, \tilde{x})$. This condition is equal to vanishing of the antiselfdual current. The antiselfdual current is nothing but the symmetry generator. The selfduality condition is second class constraint $\widetilde{\triangleright}_M = 0$ from (4.1). There are ways to make first class constraints from the second class constraints; (1) The linear combination of them as $\widetilde{\triangleright}_{\overline{M}} - \widetilde{\triangleright}_{\underline{M}} = 0$ with the 2D-dimensional indices $M = (\overline{M}, \underline{M})$ (left and right) as the dimensional reduction constraint [11]. (2) Squaring of them as $(\widetilde{\triangleright}_M)^2 = 0$ with gauging the antiselfdual modes [18]. This term is added in the Hamiltonian (2.7), (2.9) and (2.10) where Lagrange multipliers may be functions.

The Hamiltonian in curved space (2.9) is also written by

$$H = \frac{1}{2}(g_+ + g_-)\mathcal{H}_\tau + \frac{1}{2}(g_- - g_+)\mathcal{H}_\sigma + \frac{1}{2}\lambda_-(h_\tau + h_\sigma) + \frac{1}{2}\lambda_+(h_\tau - h_\sigma) \tag{4.3}$$

where $\mathcal{H}_{\tau,\sigma} = 0$ are the Virasoro constraints and $h_{\tau,\sigma} = 0$ are the selfduality constraints generating shift of the antiselfdual mode

$$\begin{cases} \mathcal{H}_\sigma &= \frac{1}{4}\overset{\square}{\triangleright}_M \eta^{MN} \overset{\square}{\triangleright}_N - \frac{1}{4}\widetilde{\triangleright}_M \eta^{MN} \widetilde{\triangleright}_N = \frac{1}{4}\left((\triangleright_{\bar{A}})^2 - (\triangleright_{\underline{A}})^2 - (\widetilde{\triangleright}_{\bar{A}})^2 + (\widetilde{\triangleright}_{\underline{A}})^2\right) \\ \mathcal{H}_\tau &= \frac{1}{4}\overset{\square}{\triangleright}_M G^{MN} \overset{\square}{\triangleright}_N + \frac{1}{4}\widetilde{\triangleright}_M G^{MN} \widetilde{\triangleright}_N = \frac{1}{4}\left((\triangleright_{\bar{A}})^2 + (\triangleright_{\underline{A}})^2 + (\widetilde{\triangleright}_{\bar{A}})^2 + (\widetilde{\triangleright}_{\underline{A}})^2\right) \\ h_\sigma &= \frac{1}{4}\widetilde{\triangleright}_M \eta^{MN} \widetilde{\triangleright}_N = \frac{1}{4}\left((\widetilde{\triangleright}_{\bar{A}})^2 - (\widetilde{\triangleright}_{\underline{A}})^2\right) \\ h_\tau &= \frac{1}{4}\widetilde{\triangleright}_M G^{MN} \widetilde{\triangleright}_N = \frac{1}{4}\left((\widetilde{\triangleright}_{\bar{A}})^2 + (\widetilde{\triangleright}_{\underline{A}})^2\right) \end{cases} \tag{4.4}$$

The Virasoro algebra is given by

$$\begin{aligned} [\mathcal{H}_\sigma(1), \mathcal{H}_\sigma(2)] &= -i(\mathcal{H}_\sigma(1) + \mathcal{H}_\sigma(2))\partial_\sigma \delta(2-1) \\ [\mathcal{H}_\sigma(1), \mathcal{H}_\tau(2)] &= -i(\mathcal{H}_\tau(1) + \mathcal{H}_\tau(2))\partial_\sigma \delta(2-1) \\ [\mathcal{H}_\tau(1), \mathcal{H}_\tau(2)] &= -i(\mathcal{H}_\sigma(1) + \mathcal{H}_\sigma(2))\partial_\sigma \delta(2-1) \end{aligned} \tag{4.5}$$

where the $\sigma$ derivatives in the canonical formalism and the usual chain rule coincide

$$\partial_\sigma \Phi|_{can} = i[\int \mathcal{H}_\sigma, \Phi] = \frac{1}{2}(\overset{\square}{\triangleright}^M - \widetilde{\triangleright}^M)\partial_M \Phi = \partial_\sigma X^M \partial_M \Phi = \partial_\sigma \Phi|_{chain} \tag{4.6}$$

No subsidiary condition on fields is required.

The bilinears of the selfduality constraints satisfy the following algebras

$$\begin{aligned} [h_\sigma(1), h_\sigma(2)] &= i(h_\sigma(1) + h_\sigma(2))\partial_\sigma \delta(2-1) \\ [h_\sigma(1), h_\tau(2)] &= i(h_\tau(1) + h_\tau(2))\partial_\sigma \delta(2-1) - i\chi\delta(1-2) \\ [h_\tau(1), h_\tau(2)] &= i(h_\sigma(1) + h_\sigma(2))\partial_\sigma \delta(2-1) \end{aligned} \tag{4.7}$$

with

$$\chi = \frac{1}{8}\widetilde{\triangleright}_N \widetilde{\triangleright}_L \overset{\square}{\triangleright}^M \partial_M G^{NL} \tag{4.8}$$

The closure of the algebra requires the section condition

$$\overset{\square}{\triangleright}{}^M \partial_M G^{NL} = 0 \tag{4.9}$$

Commutators of the Virasoro operators and the selfduality constraints are given by

$$\begin{aligned}
[\mathcal{H}_\sigma(1), h_\sigma(2)] &= -i(h_\sigma(1) + h_\sigma(2))\partial_\sigma \delta(2-1) \\
[\mathcal{H}_\sigma(1), h_\tau(2)] &= -i(h_\tau(1) + h_\tau(2))\partial_\sigma \delta(2-1) \\
[\mathcal{H}_\tau(1), h_\sigma(2)] &= i(h_\tau(1) + h_\tau(2))\partial_\sigma \delta(2-1) + i\chi_1 \delta(1-2) \\
[\mathcal{H}_\tau(1), h_\tau(2)] &= i(h_\sigma(1) + h_\sigma(2))\partial_\sigma \delta(2-1) - i\chi_2 \delta(1-2)
\end{aligned} \tag{4.10}$$

with

$$\begin{aligned}
\chi_1 &= \frac{1}{8}\left(\widetilde{\triangleright}_N \widetilde{\triangleright}_L \overset{\square}{\triangleright}_M \partial^M G^{NL}) + \overset{\square}{\triangleright}_N \overset{\square}{\triangleright}_L \widetilde{\triangleright}_M \partial^M G^{NL})\right) \\
\chi_2 &= \frac{1}{8}\left(\widetilde{\triangleright}_N \widetilde{\triangleright}_L \triangleright_A \hat{\eta}^{AB}(\nabla_B G^{NL}) + \overset{\square}{\triangleright}_N \overset{\square}{\triangleright}_L \widetilde{\triangleright}_A \hat{\eta}^{AB}(\nabla_B G^{NL})\right)
\end{aligned} \tag{4.11}$$

with $\nabla_A = E_A{}^M \partial_M$. The closure of the algebras requires the section conditions and the weaker form of the selfduality condition

$$\overset{\square}{\triangleright}{}^M (\partial_M G^{NL}) = \triangleright_A \hat{\eta}^{AB}(\nabla_B G^{NL}) = 0 \,,\; \widetilde{\triangleright}{}^M (\partial_M G^{NL}) = \widetilde{\triangleright}_A \hat{\eta}^{AB}(\nabla_B G^{NL}) = 0 \tag{4.12}$$

Let us compare the chiral description in [11]. The Virasoro constraints by the selfdual current only are given by

$$\begin{cases}
\mathcal{H}_\sigma &= \frac{1}{4}\triangleright_A \eta^{AB} \triangleright_B = \frac{1}{4}\overset{\square}{\triangleright}_M \eta^{MN} \overset{\square}{\triangleright}_N \\
\mathcal{H}_\tau &= \frac{1}{4}\triangleright_A \hat{\eta}^{AB} \triangleright_B = \frac{1}{4}\overset{\square}{\triangleright}_M G^{MN} \overset{\square}{\triangleright}_N
\end{cases} \tag{4.13}$$

They satisfy the Virasoro algebra

$$\begin{aligned}
[\mathcal{H}_\sigma(1), \mathcal{H}_\sigma(2)] &= -i(\mathcal{H}_\sigma(1) + \mathcal{H}_\sigma(2))\partial_\sigma \delta(2-1) \\
[\mathcal{H}_\sigma(1), \mathcal{H}_\tau(2)] &= -i(\mathcal{H}_\tau(1) + \mathcal{H}_\tau(2))\partial_\sigma \delta(2-1) \\
[\mathcal{H}_\tau(1), \mathcal{H}_\tau(2)] &= -i(\mathcal{H}_\sigma(1) + \mathcal{H}_\sigma(2))\partial_\sigma \delta(2-1)
\end{aligned} \tag{4.14}$$

where the $\sigma$ derivative is given by the commutator with $\mathcal{H}_\sigma$ in the canonical formalism

$$\partial_\sigma \Phi|_{can} = i[\int \mathcal{H}_\sigma, \Phi] = \frac{1}{2}\overset{\square}{\triangleright}{}^M \partial_M \Phi \tag{4.15}$$

On the other hand the usual chain rule gives

$$\partial_\sigma \Phi(X)|_{chain} = \partial_\sigma X^M \partial_M \Phi(X) \tag{4.16}$$

The equality of these differentiations requires the weaker form of the selfduality condition

$$\widetilde{\triangleright}^M \partial_M \Phi = 0 \tag{4.17}$$

In the chiral formulation the dimensional reduction constraint $\widetilde{\triangleright}_{\overline{M}} - \widetilde{\triangleright}_{\underline{M}} = 0$, which is first class, gives the following Hamiltonian

$$H = g_\tau \mathcal{H}_\tau + g_\sigma \mathcal{H}_\sigma + \lambda^{\overline{M}}(\widetilde{\triangleright}_{\overline{M}} - \widetilde{\triangleright}_{\underline{M}}) \tag{4.18}$$

leading to the worldsheet covariant string action as shown in [11]. Naively the dimensional reduction constraint supplies the antiselfdual currents.

## 5 Stringy geometry

The currents in the Lagrangian formulation are related to the currents in the Hamiltonian formulation by the following derivative relations

$$\partial_\tau \Phi = i[\int \mathcal{H}_\tau, \Phi] \ , \ \partial_\sigma \Phi = i[\int \mathcal{H}_\sigma, \Phi] \tag{5.1}$$

in the gauge $g_+ = g_- = 1$ as

$$\begin{cases} \bar{J}_+{}^{\bar{A}} = (\partial_\tau + \partial_\sigma)X^M E_M{}^{\bar{A}} = \triangleright^{\bar{A}} \\ \bar{J}_-{}^{\bar{A}} = (\partial_\tau - \partial_\sigma)X^M E_M{}^{\bar{A}} = \widetilde{\triangleright}^{\bar{A}} \\ \underline{J}_+{}^{\underline{A}} = (\partial_\tau + \partial_\sigma)X^M E_M{}^{\underline{A}} = \widetilde{\triangleright}^{\underline{A}} \\ \underline{J}_-{}^{\underline{A}} = (\partial_\tau - \partial_\sigma)X^M E_M{}^{\underline{A}} = \triangleright^{\underline{A}} \end{cases} \tag{5.2}$$

If the selfduality condition $\widetilde{\triangleright} = 0$ is imposed then $X^M$ are chiral scalars, but it is not the case in this paper.

Taking variation with respect to the Lagrange multipliers of the Lagrangian (2.16) gives Virasoro constraints $T_{++} = 0 = T_{--}$ and selfduality constraints $\bar{h} = 0 = \underline{h}$ with

$$\begin{cases} T_{++} = \frac{1}{2}(\bar{J}_+{}^{\bar{A}})^2 + \frac{1}{2}\left(\frac{g_++g_-+\lambda_+}{g_++g_-+\lambda_-}\right)^2 (\underline{J}_+{}^{\underline{A}})^2 \\ T_{--} = \frac{1}{2}\left(\frac{g_++g_-+\lambda_-}{g_++g_-+\lambda_+}\right)^2 (\bar{J}_-{}^{\bar{A}})^2 + \frac{1}{2}(\underline{J}_-{}^{\underline{A}})^2 \\ \bar{h} = \frac{1}{2}(\bar{J}_-{}^{\bar{A}})^2 \\ \underline{h} = \frac{1}{2}(\underline{J}_+{}^{\underline{A}})^2 \end{cases} \tag{5.3}$$

The T-theory space is defined by the covariant derivatives $\triangleright_A$ in Hamiltonian formulation, while the Lagrangian is described by the selfdual currents $\bar{J}_+{}^{\bar{A}}$ and $\underline{J}_-{}^{\underline{A}}$. Imposing the selfduality constraints $\bar{J}_-{}^{\bar{A}} \equiv 0 \equiv \underline{J}_+{}^{\underline{A}}$ leads to the particle (zero-mode) limit of the selfduality currents

$$\begin{aligned} \bar{J}_+{}^{\bar{A}}|_{J_{\overline{SD}}=0} &= (1+\frac{g_+}{g_-})\partial_\tau X^M E_M{}^{\bar{A}} \\ \underline{J}_-{}^{\underline{A}}|_{J_{\overline{SD}}=0} &= (1+\frac{g_-}{g_+})\partial_\tau X^M E_M{}^{\underline{A}} \ . \end{aligned} \quad (5.4)$$

The gauge invariant operators which do not vanish in the zero-mode limit are the Virasoro operators. The zero-mode limit of the Virasoro operators in the gauge $g_+ = g_- = 1$ lead to the line element of the manifest T-duality space

$$T_{++} + T_{--} = (\partial_\tau X^M E_M{}^{\bar{A}})^2 + (\partial_\tau X^M E_M{}^{\underline{A}})^2 \ \Rightarrow \ ds^2 = dX^M G_{MN} dX^N \quad (5.5)$$

and the orthogonal condition on backgrounds

$$T_{++} - T_{--} = (\partial_\tau X^M E_M{}^{\bar{A}})^2 - (\partial_\tau X^M E_M{}^{\underline{A}})^2 \ \Rightarrow \ 0 = dX^M \eta_{MN} dX^N \quad (5.6)$$

Let us examine the zero-mode limit of the curved space covariant derivative (3.9). While the vielbein is transformed with the O(D,D) matrix as (3.5) the tangent vector is transformed with the GL(2D) matrix as

$$\frac{\partial}{\partial X'^M} = \frac{\partial X^N}{\partial X'^M}\frac{\partial}{\partial X^N} \quad (5.7)$$

In order to be gauge invariant

$$E_A{}^M(X)\frac{\partial}{\partial X^M} = E'_A{}^M(X')\frac{\partial}{\partial X'^M} \quad (5.8)$$

the section condition is required

$$(\partial^M \Lambda_N)\frac{\partial}{\partial X^M} = 0 \quad (5.9)$$

It allows the O(D,D) transformation for the tangent vector as (3.8) to cancel (3.6)

$$\frac{\partial}{\partial X'^M} \approx N_M{}^N \frac{\partial}{\partial X^N} \quad (5.10)$$

Finite gauge transformations are possible if the section condition allows to make an O(D,D) matrix from a GL(2D) matrix. For consistency the section condition requires both the strong and weak section conditions for functions $\Phi(X)$, $\Psi(X)$

$$\partial^M \Psi \partial_M \Phi = \partial^M \partial_M \Psi = \partial^M \partial_M \Phi = 0 \quad (5.11)$$

It is also mentioned that the doubled coordinate $X^M$ does not satisfy the section conditions.

The zero-mode limit of the curved space one form (3.10) should be also gauge invariant

$$dX^M E_M{}^A(X) = dX'^M E'_M{}^A(X') \tag{5.12}$$

The Lagrangian version of the section condition as the zero-mode limit of (3.13) is required

$$dX^M \partial_N \Lambda_M = 0 \tag{5.13}$$

It allows the following O(D,D) transformation of the cotangent vector

$$dX'^M \approx dX^N (N^{-1})_N{}^M \tag{5.14}$$

Then the line element and the orthogonal constraint are gauge invariant under the T-duality covariant coordinate transformation

$$\begin{aligned}
ds^2 &= dX^M G_{MN}(X) dX^N = dX'^M G_{MN}(X') dX'^N \\
0 &= dX^M \eta_{MN} dX^N = dX'^M \eta_{MN} dX'^N
\end{aligned} \tag{5.15}$$

Concrete examples are shown as follows:

- Under the infinitesimal coordinate transformation

$$X^M \to X'^M = X^M - \Lambda^M , \ E'_A{}^M = E_A{}^N (\delta_N^M - \partial_N \Lambda^M + \partial^M \Lambda_N) \tag{5.16}$$

By using with the constraint the curved space one form is gauge invariant

$$dX'^M \approx dX^N (\delta_N^M - \partial_N \Lambda^M + \partial^M \Lambda_N) \Rightarrow dX^M E_M{}^A = dX'^M E'_M{}^A \tag{5.17}$$

- Under the constant O(D,D) $\ni \Xi_M{}^N$ transformation

$$X^M \to X'^M = X^N \Xi_N{}^M , \ E'_A{}^M = E_A{}^N \Xi_N{}^M \Rightarrow dX^M E_M{}^A = dX'^M E'_M{}^A \tag{5.18}$$

the curved space covariant derivative (5.8) and the curved space one form (5.12) are gauge invariant. The line element is invariant and the orthogonal condition in (5.15) is also invariant from $\Xi \eta \Xi^T = \eta$.

- The finite coordinate transformation with only $x^\mu$ dependence is examined. For convenience the O(D,D) invariant metric $\eta^{MN}$ is off-diagonal

$$X^M = (x^\mu, \tilde{x}_\mu) \to X'^M = (x'^\mu(x), \tilde{x}_\mu) , \ \eta_{MN} = \begin{pmatrix} 0 & \delta_\mu^\nu \\ \delta_\nu^\mu & 0 \end{pmatrix} \tag{5.19}$$

The coordinate transformation is a GL(2D) matrix

$$\frac{\partial X'^M}{\partial X^N} = \begin{pmatrix} a_\nu{}^\mu & 0 \\ 0 & \delta_\mu^\nu \end{pmatrix} , \ a_\nu{}^\mu = \frac{\partial x'^\mu}{\partial x^\nu} \tag{5.20}$$

By using with the section condition (5.9)

$$\frac{\partial}{\partial X'^M} \approx N_M{}^N \frac{\partial}{\partial X^N} \ , \ N_M{}^N = \frac{\partial X^N}{\partial X'^M} + \partial^N \Lambda_M \ , \ \Lambda^M = (x'^\mu(x) - x^\mu, 0) \tag{5.21}$$

the O(D,D) matrix is given as

$$N_M{}^N = \begin{pmatrix} (a^{-1})_\mu{}^\nu & 0 \\ 0 & a_\nu{}^\mu \end{pmatrix} \tag{5.22}$$

Then the cotangent vector and the vielbein are transformed by the O(D,D) matrix with the condition (5.13) as

$$dX'^M \approx dX^N (N^{-1})_N{}^M \ , \ E'_M{}^A = N_M{}^N E_N{}^A \tag{5.23}$$

As a result both the line element and the orthogonal condition are invariant under the finite gauge transformation.

It is also denoted that the contravariant vector $V^M(X)$ is transformed by the new Lie derivative (3.1)

$$\mathcal{L} V^M = \Lambda^N \partial_N V^M - V^N (\partial_N \Lambda^M - \partial^M \Lambda_N) \tag{5.24}$$

and the covariant vector is its inversely transformed.

The relation to the Buscher's transformation [23, 24] is also mentioned by taking AdS space as a simple example. The Lagrangian for a string in the D-dimensional AdS space is given by

$$L = \tfrac{1}{2} \partial_+ x^\mu G_{\mu\nu} \partial_- x^\nu \ , \ G_{\mu\nu} = \frac{1}{(x^0)^2} \tag{5.25}$$

with $x^\mu = (x^{\hat\mu}, x^0)$. Under the Buscher's T-duality transformation with respect to D−1 coordinates $x^{\hat\mu}$ both the coordinates and the gravitational metric are transformed,

$$x^{\hat\mu} \to \tilde{x}_{\hat\mu} \ , \ G_{\hat\mu\hat\nu} \to \tilde{G}^{\hat\mu\hat\nu} \tag{5.26}$$

with the selfduality condition

$$\partial_\tau x^{\hat\mu} G_{\hat\mu\hat\nu} = \partial_\sigma \tilde{x}_{\hat\nu} \ , \ \partial_\sigma x^{\hat\mu} G_{\hat\mu\hat\nu} = \partial_\tau \tilde{x}_{\hat\nu} \tag{5.27}$$

The Lagrangian becomes

$$L = \tfrac{1}{2} \partial_+ \tilde{x}_{\hat\mu} \tilde{G}^{\hat\mu\hat\nu} \partial_- \tilde{x}_{\hat\nu} + \tfrac{1}{2} \partial_+ x^0 G_{00} \partial_- x^0 \ , \ \tilde{G}^{\hat\mu\hat\nu} = (x^0)^2 \tag{5.28}$$

The coordinate redefinition $x^0 \to \tilde{x}_0 = 1/x^0$ gives the final expression

$$L = \tfrac{1}{2} \partial_+ \tilde{x}_\mu \tilde{G}^{\mu\nu} \partial_- \tilde{x}_\nu \ , \ \tilde{G}^{\mu\nu} = \frac{1}{(\tilde{x}_0)^2} \tag{5.29}$$

On the other hand we begin by the T-theory Lagrangian in the conformal gauge

$$L = \tfrac{1}{2}\partial_+ x^M G_{MN} \partial_- x^N \,, \quad G_{MN} = \begin{pmatrix} \frac{1}{(x^0)^2} & 0 \\ 0 & \frac{1}{(\tilde{x}_0)^2} \end{pmatrix} \quad (5.30)$$

where the O(D,D) invariant metric is offdiagonal (5.19). The Buscher's T-duality transformation is a discrete O(D,D) rotation $\Xi$

$$X^M = (x^{\hat{\mu}}, x^0, \tilde{x}_{\hat{\mu}}, \tilde{x}_0) \,, \quad \Xi_M{}^N = \begin{pmatrix} & & \delta^{\hat{\nu}}_{\hat{\mu}} & \\ & 1 & & \\ \delta^{\hat{\mu}}_{\hat{\nu}} & & & \\ & & & 1 \end{pmatrix} \quad (5.31)$$

which is inserted in the Lagrangian trivially as $\partial_+ X \Xi \Xi^T G \Xi \Xi^T \partial_- X$. The Lagrangians (5.25) and (5.29) are different sections, the $x^\mu$ coordinate section and the $\tilde{x}_\mu$ coordinate section respectively. The O(D,D) condition implies

$$x^0 \, \tilde{x}_0 = 1 \quad (5.32)$$

# 6 Conclusions

The string action in curved space with manifest T-duality is presented. The gravitational field is O(D,D) gauge field and the gauge invariance of the Lagrangian is shown. Its gauge invariance requires the Lagrangian version of the section condition and the weaker form of the selfduality condition. Finally the line element and the orthogonal condition in the manifest T-duality space are presented and their gauge invariances are also shown.

This formulation will be useful for quantum computation to explore T-duality symmetric situations.

# Acknowledgements

M.H. is most grateful to Kiyoshi Kamimura for fruitful discussions. We lso would like to thank Di Wang for useful discussions. M.H. would like to thank the Simons Center for Geometry and Physics for hospitality during "the 2018 Summer Simons workshop in Mathematics and Physics" where this work has been developed. W.S. is supported by NSF grant PHY-1620628.

# A   Indices

Indices are summarized.

$$b\text{(egining)} : \text{flat} \cdots A, B, \cdots$$
$$m\text{(iddle)} : \text{curved} \cdots M, N, \cdots$$

$$\textit{UPPER CASE} : \text{spacetime} \cdots M, N, \cdots$$
$$\textit{lower case} : \text{worldvolume} \cdots m, n, \cdots$$

$$\bar{B}\text{arred} : \text{left-handed} \cdots \bar{A}, \bar{B}, \cdots, \overline{M}, \overline{N}, \cdots$$
$$\underline{U}\text{nderlined} : \text{right-handed} \cdots \underline{A}, \underline{B}, \cdots, \underline{M}, \underline{N}, \cdots$$